\documentclass[aps,10pt,prb,twocolumn,letterpaper,superscriptaddress,floatfix]{revtex4-2}
\usepackage{graphicx} % Include figure files
\usepackage{dcolumn} % Align table columns on decimal point
\usepackage{bm} % bold math
\usepackage{siunitx} % SI units
\usepackage{physics} % Physics symbols
\usepackage{xcolor}
\definecolor{PRLblue}{rgb}{0.18,0.18,0.57}
\usepackage[colorlinks=true,allcolors=blue]{hyperref} % add hypertext capabilities
\usepackage{amssymb}

\usepackage{fancyhdr}
\usepackage[version=4]{mhchem}

\begin{document}

%\preprint{APS/123-QED}

\title{Copper migration and surface oxidation of \ce{Cu_xBi2Se3} in ambient pressure environments}

% Force line breaks with \\

\author{Adam L. Gross}
\affiliation{Department of Physics and Astronomy, University of California, Davis, CA 95616, USA}
\author{Lorenz Falling}
\affiliation{Advanced Light Source, Lawrence Berkeley National Lab, Berkeley, 94720, USA}
\author{Matthew C. Staab}
\affiliation{Department of Physics and Astronomy, University of California, Davis, CA 95616, USA}
\author{Metzli I. Montero}
\affiliation{Department of Physics and Astronomy, University of California, Davis, CA 95616, USA}
\author{Rahim R. Ullah}
\affiliation{Department of Physics and Astronomy, University of California, Davis, CA 95616, USA}
\author{David M. Nisson}
\affiliation{Department of Physics and Astronomy, University of California, Davis, CA 95616, USA}
\author{Peter Klavins}
\affiliation{Department of Physics and Astronomy, University of California, Davis, CA 95616, USA}
\author{Kristie J. Koski}
\affiliation{Department of Chemistry, University of California, Davis, CA 95616, USA}
\author{Nicholas J. Curro}
\affiliation{Department of Physics and Astronomy, University of California, Davis, CA 95616, USA}
\author{Valentin Taufour}
\affiliation{Department of Physics and Astronomy, University of California, Davis, CA 95616, USA}
\author{Slavomir Nemsak}
\affiliation{Department of Physics and Astronomy, University of California, Davis, CA 95616, USA}
\affiliation{Advanced Light Source, Lawrence Berkeley National Lab, Berkeley, 94720, USA}
\author{Inna M. Vishik}%
\affiliation{Department of Physics and Astronomy, University of California, Davis, CA 95616, USA}

\date{\today}% It is always \today, today,

\begin{abstract}

Chemical modifications such as intercalation can be used to modify surface properties or to further functionalize the surface states of topological insulators. Using ambient pressure X-ray photoelectron spectroscopy, we report copper migration in \ce{Cu_xBi2Se3}, which occurs on a timescale of hours to days after initial surface cleaving. The increase in near-surface copper proceeds along with the oxidation of the sample surface and large changes in the selenium content. These complex changes are further modelled with core-level spectroscopy simulations, which suggest a composition gradient near the surface which develops with oxygen exposure. Our results shed light on a new phenomenon that must be considered for intercalated topological insulators—and intercalated materials in general—that surface chemical composition can change when specimens are exposed to ambient conditions. 

\end{abstract}

\maketitle

\subsection{Introduction}

Topological insulators (TIs) have attracted significant interest from the quantum materials community for their surface electronic structures, notably surface Dirac fermions with spin-momentum locking that are robust to disorder and defects \cite{Ando2013,Hasan2010,Wehling2014,Keimer2017}. Chemical intercalation of TIs (i.e., the insertion of atoms or molecules between quintuple layers) can further modify the material properties \cite{Paraskevopoulos1988,Bludska2010,Choi2011, Shruti2015,Mazumder2018,Yonezawa2018,Wang2020} or create additional electronic phenomena, such as superconductivity with Cu intercalation ($T_{c,\text{max}} \sim~ \SI{3.5}{\kelvin}$) \cite{Wray2010,Hor2010,Sasaki2011,Kriener2011,Das2013,Yonezawa2016,Yonezawa2018}. The Cu-intercalated TI \ce{Bi2Se3} (\ce{Cu_xBi2Se3}) is a known superconductor, with $T_{c,\text{max}}$ and the superconducting shielding fraction strongly depending on the intercalated Cu content and preparation method \cite{Hor2010, Sasaki2011, Kriener2011, Schneeloch2015}. Controlling the Cu content and its variation during and after synthesis is thus important for realizing ideal superconducting properties in \ce{Cu_xBi2Se3}.

Intercalated materials may be exposed to air during sample preparation or real world usage, such as in the operation of devices, which may affect composition. For instance, the intercalate guest can diffuse within battery cells left at a fixed potential, and it is known that intercalants can diffuse within and between layers in various compounds. \cite{Whittingham1978,Ryabishchenkova2015, Ye2021}. The effect of ambient conditions on intercalant chemistry is often not fully understood due to experimental difficulties, despite its practical importance.

In this work, we use \ce{Cu_xBi2Se3} as a representative material to establish the effects of controlled ambient environments on evolving chemistry in the near-surface region in intercalated TIs. The near-surface Cu composition was found to increase under both controlled \ce{O2} dosing and ambient air exposure, coincident with the formation and growth of an oxide layer that is strongest in full atmosphere. Lastly, core-electron spectroscopy simulations show that our XPS observations are consistent with a sample developing vertical gradient distributions of \ce{Cu} and \ce{Se} upon exposure to a controlled \ce{O2} environment.

\begin{figure*}[t]
\includegraphics[width=1.0\textwidth]{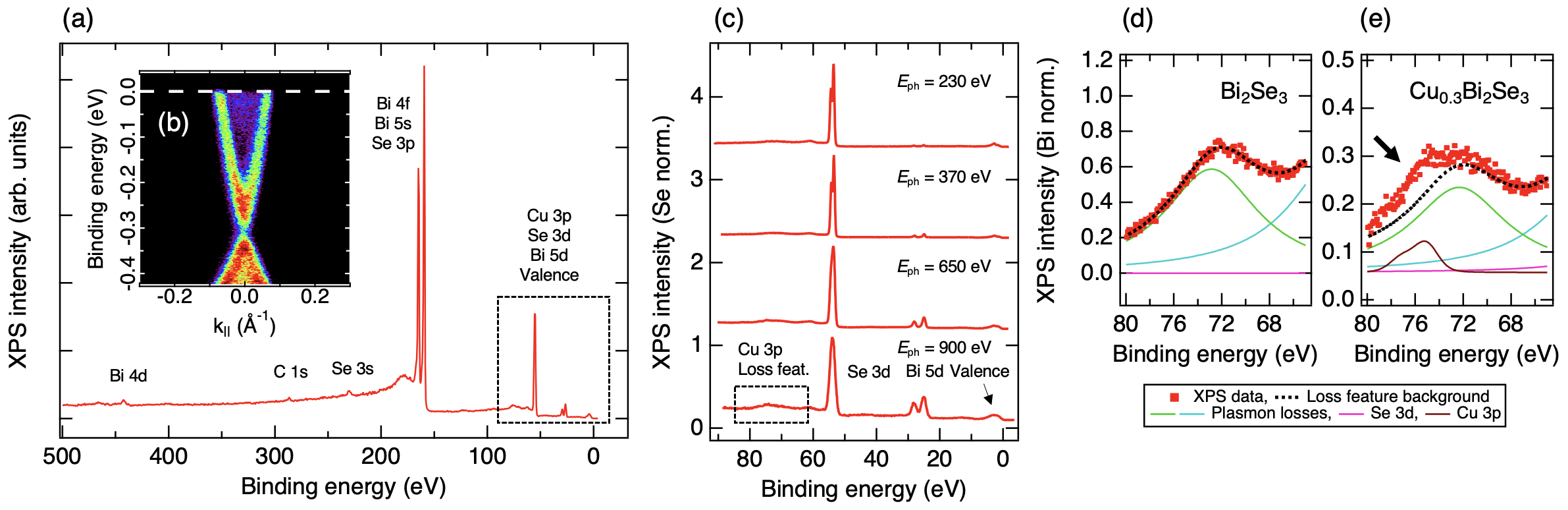}
\caption{XPS and ARPES characterizations of freshly cleaved \ce{Cu_{0.15}Bi2Se3} and \ce{Cu_{0.3}Bi2Se3}. (a) \ce{Cu_{0.15}Bi2Se3} survey spectrum taken with $E_{ph} = \SI{650}{\electronvolt}$ in vacuum. Cu, Bi, and Se core levels are labeled, with the dashed box indicating the shallow core levels analyzed in this work. The adventitious C 1s peak is also labeled. (b) ARPES spectrum of the topological surface state in \ce{Cu_{0.15}Bi2Se3}. (c) Shallow core levels in the dashed box taken with $E_{ph} = \SI{230}{\electronvolt}$, \SI{370}{\electronvolt}, \SI{650}{\electronvolt}, and \SI{900}{\electronvolt} in vacuum. For best comparison, spectra are shown normalized to their Se 3d peaks and offset for clarity. (d) Illustration of the loss feature background in a \ce{Bi2Se3} sample, with the individual plasmon resonances indicated in green and cyan. (e) Corresponding spectrum in \ce{Cu_{0.3}Bi2Se3} showing the Cu 3p peaks (brown curve) and loss feature background after dosing \SI{0.133}{\milli\bar} \ce{O2}. The black arrow indicates the additional spectral weight from the Cu. (d) and (e) were taken with $E_{ph} = \SI{370}{\electronvolt}$.}
\label{fig_1}
\end{figure*}

\subsection{Materials and methods}
Synchrotron ambient pressure XPS (AP-XPS) experiments were performed at the Advanced Light Source Beamline 9.3.2 in a photon energy range of $E_{ph} = 230-\SI{900}{\electronvolt}$, with a spot size of $d\sim\SI{1}{\milli\meter\squared}$. The core level spectra were collected with a Scienta R4000 HiPP electron analyzer with differential pumping, allowing the sample to remain at ambient pressures during data acquisition \cite{Grass2010}. The analysis environment was initially at high vacuum ($\sim\SI{1e-7}{\milli\bar}$), and then with \ce{O2} at ambient pressure (\SI{0.133}{\milli\bar}). The measured partial pressure of \ce{O2} is equivalent to the total ambient pressure in the AP-XPS experiment. Longer timescale experiments were performed using a Kratos AXIS Supra$^+$ with a monochromated, unpolarized $E_{ph} = \SI{1486.6}{\electronvolt}$ Al-K$\alpha$ source. Samples were cleaved \textit{in situ} with the top post method in the AP-XPS experiment, and cleaved \textit{ex situ} with Scotch tape in the Kratos XPS experiment, keeping air exposure under several minutes prior to the initial measurements. ARPES spectra were collected at the Advanced Light Source Beamline 4.0.3 (MERLIN) with $E_{ph} = \SI{39.2}{\electronvolt}$. Core level peaks were identified using binding energy reference values \cite{X_ray_handbook}.

\ce{Bi2Se3} samples were grown using the Bridgman method, and were then intercalated with Cu to form \ce{Cu_xBi2Se3} crystals using a solution-based process \cite{Koski2012}. Due to the variation of Cu and Se content highlighted in our study, our convention is to use the nominal bulk stoichiometries after synthesis (\ce{Cu_{0.15}Bi2Se3} and \ce{Cu_{0.3}Bi2Se3}) to distinguish between samples in our analysis. No indications of superconductivity were found in these samples, which depends on the details of the synthesis and the intercalation \cite{Schneeloch2015, Yu2019} (see Supplemental Material for further details).

\subsection{Results}

Figure \ref{fig_1} provides initial XPS and ARPES characterizations of \ce{Cu_{0.15}Bi2Se3} and \ce{Cu_{0.3}Bi2Se3} at several different photon energies. Fig. \ref{fig_1}(a) shows a survey spectrum of the core levels accessible with $E_{ph} = \SI{650}{\electronvolt}$ in the AP-XPS experiment. The survey confirms the quality of the \textit{in situ} cleaved sample, showing peaks for Cu, Bi, and Se. Fig. \ref{fig_1}(b) shows an ARPES spectrum on an \textit{in situ} cleaved sample, clearly showing the topological surface state and Dirac point at $E_B\sim\SI{0.3}{\electronvolt}$, indicating maintenance of crystallinity and topological electronic features after Cu intercalation. Fig. \ref{fig_1}(c) shows the three core levels present in the shallow binding energy region in Fig. \ref{fig_1}(a): Cu 3p, Se 3d, and Bi 5d, none of which show initial oxidation or hydroxylation. This set of three shallow core levels is sufficient to determine the chemical composition of \ce{Cu_{0.15}Bi2Se3} and \ce{Cu_{0.3}Bi2Se3} throughout the oxidation process; the relative peak intensities are proportional to the elemental composition after correcting for the relative sensitivity factor (RSF) for each photon energy \cite{Yeh1985, Shard2020}. To quantify the Cu content, we monitor the binding energy region around $E_B\sim \SI{76}{\electronvolt}$ where the Cu 3p doublet is present. All Cu 3p data are measured above the loss feature-containing background (Figs. \ref{fig_1}(d,e)).

To accurately quantify the Cu content in Fig. \ref{fig_1}, which is present on top of a relatively large background, we utilize a background correction procedure incorporating the electron energy loss features in the \ce{Cu_{0.3}Bi2Se3} XPS spectrum. The loss features near $E_B\sim \SI{72}{\electronvolt}$ (green and cyan curves in Figs. \ref{fig_1}(d,e)) originate from bulk plasmon resonances in \ce{Bi2Se3} \cite{Nascimento1999}. Since this background is also present in a \ce{Bi2Se3} reference sample from the same batch (Fig. \ref{fig_1}(d)), we use the \ce{Bi2Se3} spectrum from a newly cleaved sample as a reference for fitting to the plasmon losses in the \ce{Cu_{0.3}Bi2Se3} spectrum (Fig. \ref{fig_1}(e)). After fitting Voigt peaks to the Se 3d peaks and the loss features in the reference spectrum, the result is a background containing the loss features (dashed curve). Fitting this background to the \ce{Cu_{0.3}Bi2Se3} spectrum allows us to isolate the Cu 3p signal intensity, shown as a separate Voigt doublet (brown curve). The areas of the fitted Cu, Se, and Bi peaks then correspond to the total XPS intensity for these elements. Additional information on the Cu quantification, RSF assumptions, and plasmon loss feature correction can be found in the Supplemental Material.

Figure \ref{fig_2} shows the main results of the AP-XPS experiment, which track the evolution of the Cu and Se composition from an initial condition with no \ce{O2} dosing, and later times in \SI{0.133}{\milli\bar} \ce{O2} to simulate ambient atmosphere. The chemical compositions are expressed in terms of the atomic fractions of Cu/Bi and Se/Bi for each time and photon energy, with $x = 2\times\text{Cu/Bi}$. To determine the atomic fractions, the XPS intensities determined from the peak fitting were weighed by the RSF for each element and photon energy (see Supplemental Material). The atomic fractions are normalized to Bi since Bi changes relatively little compared to Cu and Se in our experiments. 

Over the course of the four day experiment, the Cu fraction $x$ increases from $x = 0.13$ to $x = 0.40$ after dosing \ce{O2} for the $E_{ph} = \SI{900}{\electronvolt}$ data, with smaller changes seen at lower, more surface sensitive $E_{ph}$ (Fig. \ref{fig_2}(a)). The growth of the Cu/Bi ratio is consistent with Cu migrating vertically to the probed-surface region of the sample, since the only additional Cu present is deeper within the sample, with no external Cu deposition. This is best seen when expressed versus the inelastic mean free path (IMFP) for each photon energy in Fig. \ref{fig_2}(b), which provides a length scale for the observed Cu migration. The measured depth ($3\times$IMFP) at each photon energy includes depths below one IMFP, which still contribute to the XPS measurement. A similar plot of the Se/Bi fraction is shown in Fig. \ref{fig_2}(c), showing a steady decrease in Se/Bi at all $E_{ph}$ after dosing \ce{O2}, decreasing from the ideal Se/Bi = 3/2 fraction of \ce{Bi2Se3}. There are large differences in the initial and final Se/Bi fractions for the different $E_{ph}$, implying a large variation in the Se distribution, with the bulk rich in Se and the surface deficient in Se (Fig. \ref{fig_2}(d)). Between one and four days in the AP-XPS chamber, the changes to Cu/Bi and Se/Bi are less pronounced than those that happen immediately after dosing \ce{O2}. Similar trends have also been observed in another AP-XPS experiment in a mixed \ce{O2}/\ce{H2O} environment (see Supplemental Material).

Figure \ref{fig_3} shows the initial and final Bi 5d peaks for all $E_{ph}$ before and after dosing \ce{O2}, quantifying the growth of oxides. The initial state peaks are well separated, showing little to no signs of Bi oxidation (Fig. \ref{fig_3}(a)). After four days in the AP-XPS chamber, chemically-shifted peaks are present for $E_{ph} = \SI{650}{\electronvolt}, \SI{370}{\electronvolt}$, and $\SI{230}{\electronvolt}$ (Fig. \ref{fig_3}(b), dashed colored lines). These changes to the Bi levels directly show the formation of \ce{Bi2O3} at the surface, which is most dominant at lower $E_{ph}$. The noise in the initial $E_{ph} = \SI{230}{\electronvolt}$ spectrum (blue, left) is due to the low photon flux and low photoionization cross-section of Bi at this particular $E_{ph}$, which was ameliorated in the final state (blue, right) by extending the measurement time. The large width of the oxide components in the Bi 5d spectra suggest a convolution of several electrostatically (i.e., band bending) or chemically distinctive states, which are most pronounced in the spectrum measured with highest surface sensitivity ($E_{ph} = \SI{230}{\electronvolt}$). Fig. \ref{fig_3}(c) quantifies the growth of the \ce{Bi2O3} peaks relative to the total Bi peak intensity, showing that oxidation starts immediately  after dosing \ce{O2}, with a larger oxide fraction seen with the more surface-sensitive $E_{ph} = \SI{370}{\electronvolt}$ (compared to $E_{ph} = \SI{650}{\electronvolt})$. Interestingly, most of the oxidation process occurs within the first day and only little change is observed thereafter, as is the case with the Cu and Se changes.

\begin{figure}[h]
\includegraphics[width=0.5\textwidth]{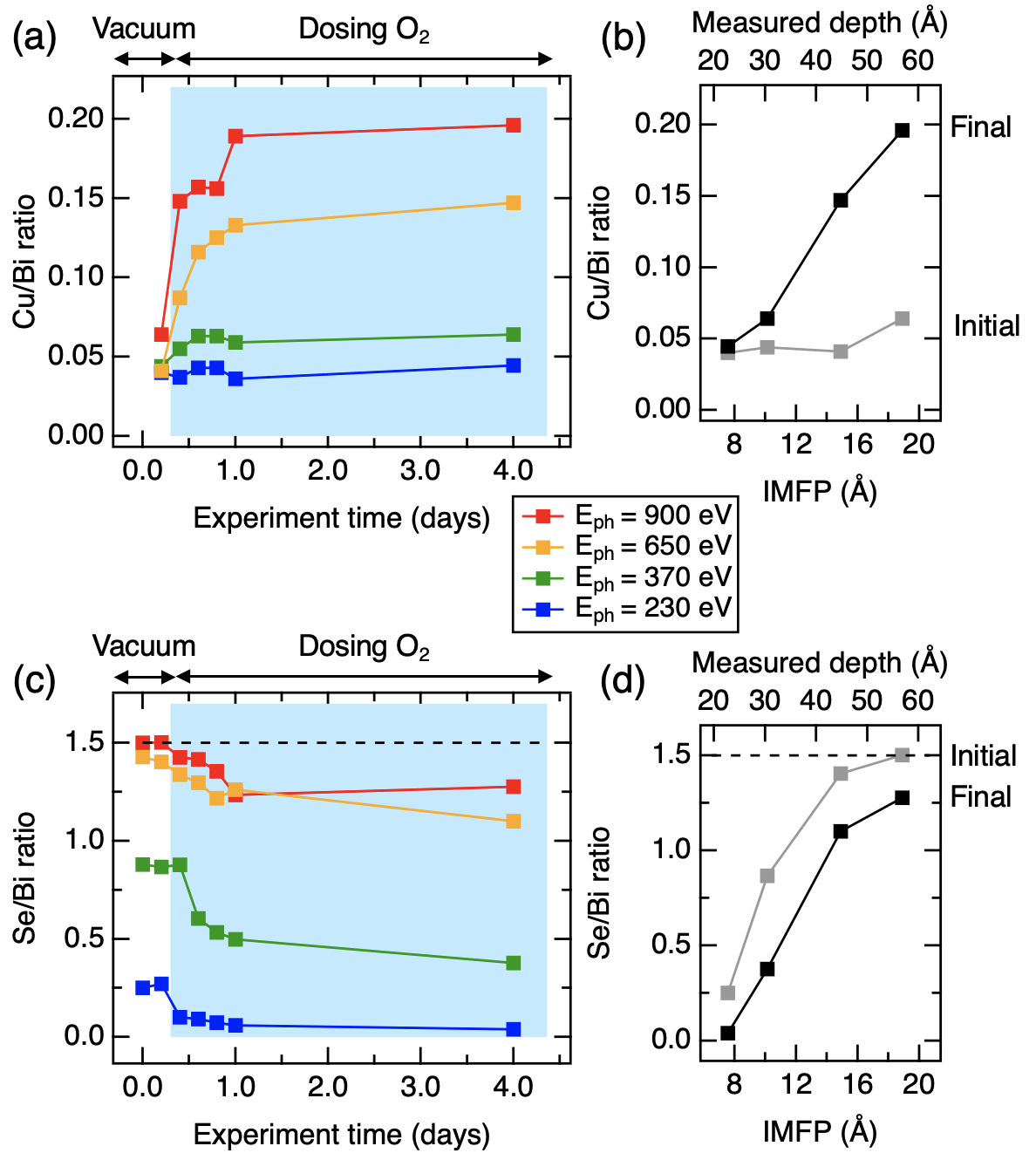}
\caption{Evolution of Cu and Se composition in \ce{Cu_{0.3}Bi2Se3} in the AP-XPS experiment. (a) Cu/Bi atomic ratios for four different photon energies, initially in vacuum and after in \SI{0.133}{\milli\bar} \ce{O2}. (b) Initial and final Cu/Bi ratios expressed in terms of the IMFP and the measured depth ($3\times$IMFP) for each $E_{ph}$ as determined from (a). Initial: in vacuum after cleaving; Final: in \ce{O2} for 4 days. (c) Se/Bi atomic ratios for the same photon energies and conditions. Dashed line at Se/Bi $= 1.5$ indicates the ideal Se/Bi = $3/2$ ratio. (d) Initial and final Se/Bi ratios from (c) expressed in terms of IMFP and measured depth.}
\label{fig_2}
\end{figure}

To gain insight into changes in the \ce{Cu_{0.15}Bi2Se3} composition over longer timescales ($t > \SI{20}{\hour}$), we continue by performing a different set of lab-based ex-situ XPS measurements after exposing a \ce{Cu_{0.15}Bi2Se3} sample to full atmosphere in air, with increasing periods of air exposure. The loss feature-corrected spectra for Cu 3p are shown in Fig. \ref{fig_4}(a), with large changes in the Cu/Bi fraction after exposure to air for multiple days (Fig. \ref{fig_4}(b)). In Fig. \ref{fig_4}(c), we again see the oxidation of the surface, with the Bi oxides showing stronger growth over several weeks. The evolution of \ce{Bi2O3} in Fig. \ref{fig_4}(d) shows an initial jump after exposure to air for 0.9 days, and continues to grow afterward. Oxidation of Se is also seen (see Supplemental Material), and we note that this sample had a smaller initial Cu concentration than the sample measured with AP-XPS.

\begin{figure}[h]
\includegraphics[width=0.5\textwidth]{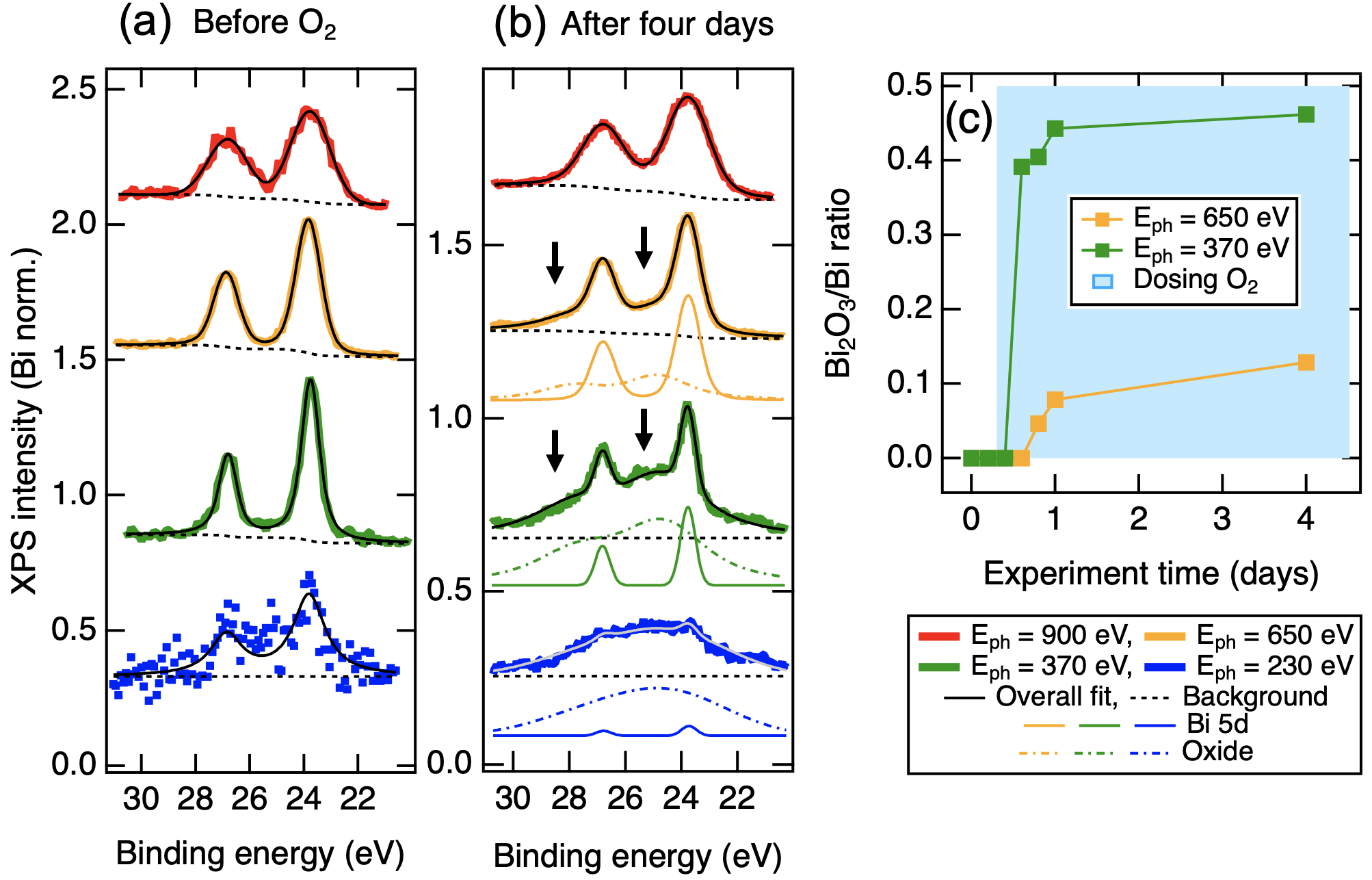}
\caption{Oxidation of \ce{Bi} in \ce{Cu_{0.3}Bi2Se3} in the AP-XPS experiment. (a) Bi 5d levels before dosing \ce{O2} and (b) after four days of oxidation in \SI{0.133}{\milli\bar} \ce{O2}. Solid colored lines indicate the Bi 5d components of the overall fits, and the dashed colored lines indicate the \ce{Bi2O3} contributions. (c) \ce{Bi2O3}/Bi ratio over the course of the entire experiment for two photon energies.}
\label{fig_3}
\end{figure}

\begin{figure}[t]
\includegraphics[width=0.45\textwidth]{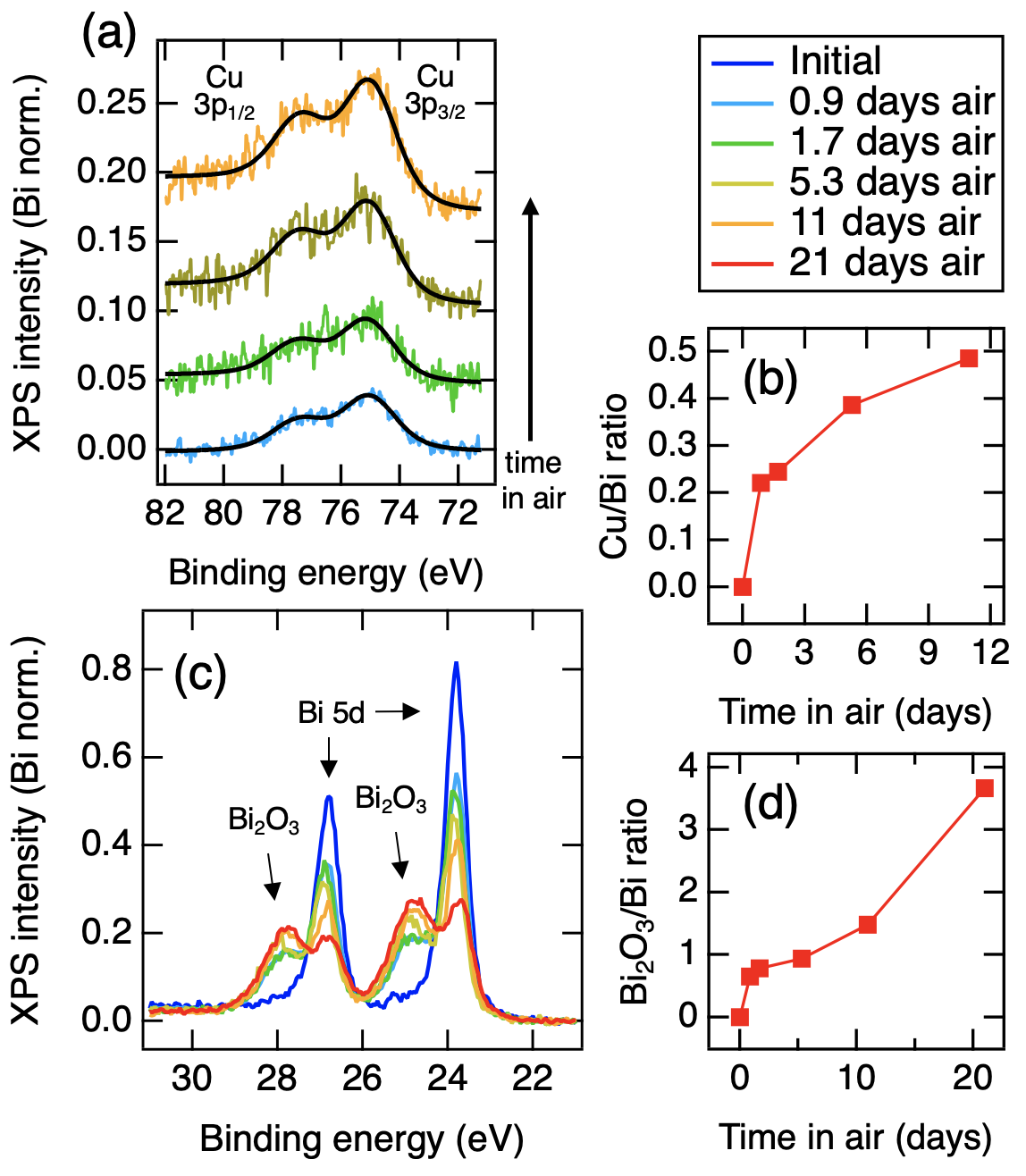}
\caption{Growth of Cu 3p and Bi oxidation at longer ($t > \SI{20}{\hour}$) timescales in \ce{Cu_{0.15}Bi2Se3}. (a) Increasing Cu 3p signal over 11 days of air exposure with $E_{ph} = \SI{1486.6}{\electronvolt}$. All Cu 3p spectra are loss feature-corrected and normalized to their total Bi 5d intensities, including oxide peaks. (b) Cu/Bi atomic ratio after air exposure. (c) Evolution of Bi 5d peaks over time, showing the appearance of \ce{Bi2O3} peaks. (d) Growth of \ce{Bi2O3}/\ce{Bi} over many days.}
\label{fig_4}
\end{figure}

\subsection{Discussion}

Our interpretation of present data is guided by prior studies of pristine \ce{Bi2Se3} exposed to air and controlled ambient environments. The effects of ambient conditions in regular \ce{Bi2Se3} have been studied in the following ways with the following conclusions. Exposure to ambient environments has been shown to alter measured angle-resolved photoemission spectroscopy (ARPES) spectra and the surface composition in \ce{Bi2Se3} \cite{Benia2011,Chen2012,Green2016,Thomas2015}. The topological surfaces states can be modified by forming 2D quantum well states in ambient conditions \cite{Chen2012}, and can present band-bending and controlled charge-doping after dosing \ce{H2O} \cite{Benia2011} and UV irradiation \cite{Sakamoto2021}. In most cases, the robust topological surface states are still present in \ce{Bi2Se3} despite air exposure and oxidation \cite{Chen2012,Yang2020}. The thinness of the oxide layer and the robustness of the topological surface states are common themes in \ce{Bi2Se3}, including the persistence of surface states in intercalated samples \cite{Tanaka2012,Ye2021}, with some works not reporting any surface reactivity \cite{Yashina2013,Atuchin2011}. Still, the surface chemistry in \ce{Bi2Se3} remains an open question, particularly after intercalation.

To begin the discussion of our results, we first turn to the observations seen in the AP-XPS experiment. The main result in Fig. \ref{fig_2} shows that the increase in Cu 3p peak intensity over the course of the experiment is coincident with the introduction of \ce{O2} gas, showing that Cu migrates to the surface region during measurement. This behavior has not been previously reported or quantified in a topological insulator with XPS, although some indications of Cu near the surface have been reported with STM imaging after cleaving \cite{Hor2010}.

The probing depth in XPS is mainly limited by the inelastic mean free path (IMFP) of escaped electrons at each $E_{ph}$, so measurement with several different $E_{ph}$ to vary IMFP allows one to obtain a depth profile of the elements in the sample. Between the photon energy range $E_{ph} = 230-\SI{900}{\electronvolt}$ in \ce{Cu_{0.3}Bi2Se3}, the IMFP of Cu 3p photoelectrons ranges from $7.6-\SI{18.9}{\angstrom}$, calculated in QUASES using the TPP2M algorithm \cite{Tanuma1994}. 

Measured XPS intensities generally follow an exponential form for attenuation \cite{Tougaard2021}: \begin{equation}\label{eqn1}
dI=I_0 \cdot X(z) \cdot e^{-z/\lambda\cos(\theta)}dz,
\end{equation}

with a total emitted photoelectron intensity $I_0$, vertical depth $z$, atomic fraction $X_A(z)$ at each depth $z$, the IMFP $\lambda$, and the photoelectron collection angle $\theta$ from the surface normal. From Eq. \ref{eqn1}, the measured intensity is $I \approx 0.95I_0$ within three IMFPs below the surface, providing an upper limit for the probing depth. Fig. \ref{fig_2}(b) shows the initial and final distributions of Cu/Bi as a function of IMFP and the measured depth ($3\lambda$) in the AP-XPS experiment (black and gray solid lines), showing the growth of Cu from deeper within the sample. The greatest Cu increase is at a measured IMFP of \SI{18.9}{\angstrom}, which corresponds to Cu migrating into the top \SI{6}{\nano\meter} surface region of the sample.

Another notable observation is the decrease in Se content relative to Bi in Fig. \ref{fig_2}(c) over the course of the AP-XPS experiment. Initially after cleaving (that is, prior to the first data points in Fig. \ref{fig_2}(c)), the Se 3d intensity is already reduced, which continues to drop over the course of the experiment from the ideal $\ce{Se}/\ce{Bi}= 3/2$. The initial distribution of Se when expressed versus IMFP in Fig. \ref{fig_2}(d) shows that the uppermost \SI{2}{\nano\meter} (2 quintuple layers) is deficient in Se, while the uppermost \SI{6}{\nano\meter} (6 quintuple layers) is richer in Se and representative of bulk stoichiometric \ce{Bi2Se3}. After four days in \ce{O2}, Se/Bi decreases at all depths. This observation of a long Se gradient with minimal Se at the surface is consistent with the well known volatility of Se in \ce{Bi2Se3}. \ce{Bi2Se3} generally has selenium vacancies which make samples naturally $n$-type without further chemical compensation. \cite{Chen2012,Biswas2015,Benia2011,Bianchi2010,Kong2011,Tumelero2016,Wang2011,Hou2019,Gross2021}. However, our results indicate that these selenium vacancies may be more concentrated near the surface, and the deeper bulk is closer to nominal stoichiometry. The Fermi level $E_F$ in the ARPES spectrum in Fig. \ref{fig_1}(b) intersects the bulk conduction band, confirming that this specimen is $n$-type at the surface region, with an IMFP of \SI{4.9}{\angstrom} for $E_{ph} = \SI{39.2}{\electronvolt}$ calculated in QUASES. Cu doping is also known to shift the chemical potential further into the bulk conduction band \cite{Wang2011}.

A prior XPS/AFM study has observed small Bi islands that appear within one hour after cleaving \cite{Green2016}, which is also consistent with decreasing Se/Bi in Fig. \ref{fig_2}(c), suggesting that some Bi migration could also be occurring along with Cu. However, the larger increases in Cu/Bi we see suggest that any Bi migration would be very small, below the sensitivity of a standard XPS instrument \cite{Green2016} and indistinguishable from Bi oxides at the surface. 

\begin{figure}[h]
\includegraphics[width=0.45\textwidth]{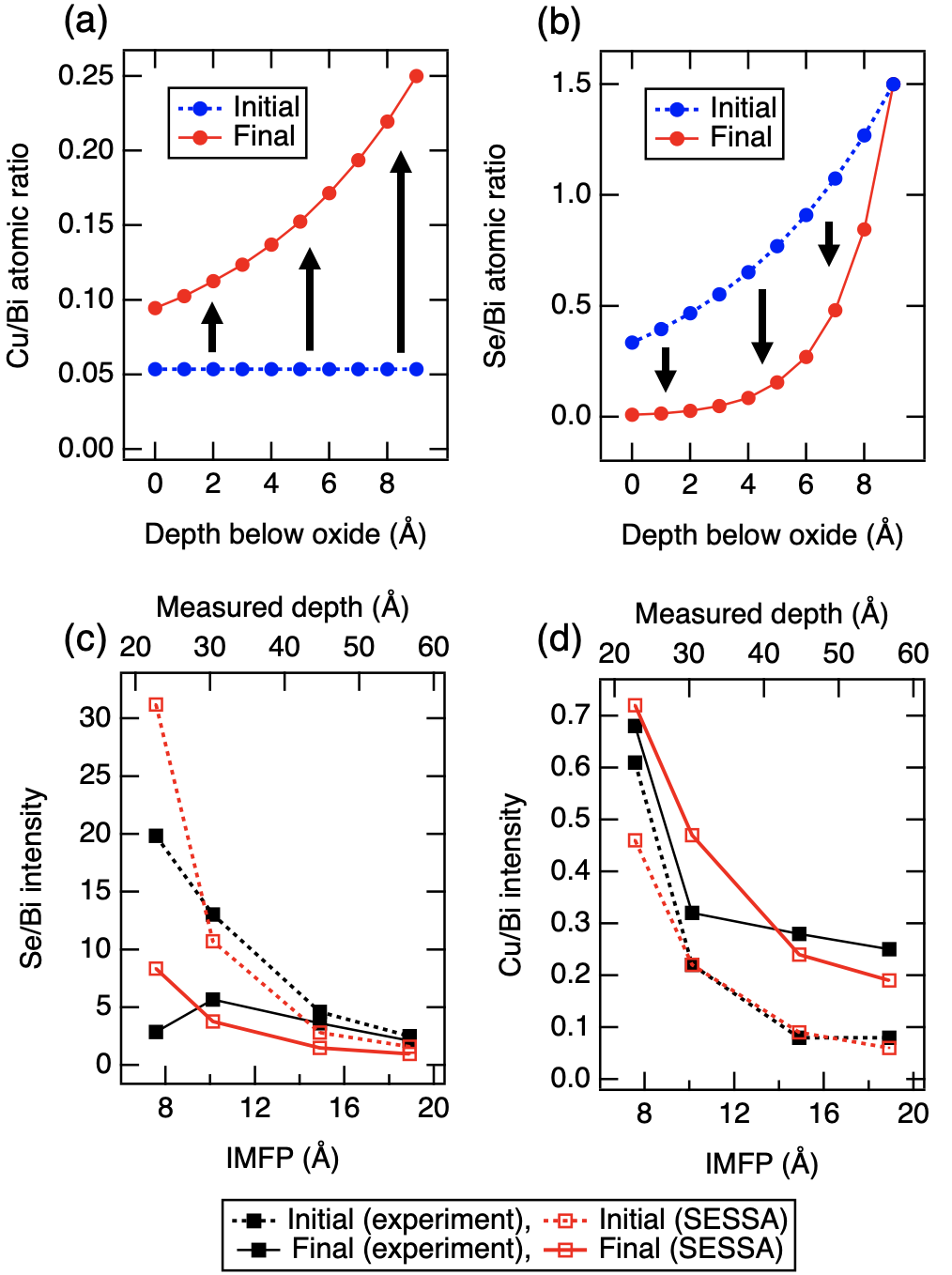}
\caption{SESSA modelling of surface compositional gradients in \ce{Cu_xBi2Se_y}. (a) Proposed Cu distributions for the initial state (before dosing \ce{O2}) and the final state (after dosing \ce{O2} for four days). The initial Cu distribution is uniform, and the final Cu distribution has developed a Cu gradient, including a \SI{2}{\angstrom} \ce{Bi2O3} overlayer. Black arrows indicate the passage of time. (b) Proposed Se gradients before and after dosing \ce{O2}. The final Se distribution has a steeper gradient when nearing the surface, with a \SI{2}{\angstrom} \ce{Bi2O3} overlayer. (c) Comparison of the AP-XPS Cu/Bi intensity ratios (black squares) to SESSA simulated ratios (red squares) for the initial and final states. (d) AP-XPS Se/Bi intensity ratios compared to SESSA values.}
\label{fig_5}
\end{figure}

While Cu migration can be seen on shorter timescales, the process continues and is more easily seen at the longer timescales ($t > \SI{25}{\hour}$) in Fig. \ref{fig_4}(a). On these timescales the oxidation of the near-surface Bi is evident with new oxide peaks, consistent with prior work \cite{Green2016,Kong2011} and the oxide peaks we see during the AP-XPS experiment in Fig. \ref{fig_3}. The oxidation is stronger and continues for longer in full atmosphere than in the AP-XPS experiment ($P=\SI{1.3e-4}{}$ atm), even when considering the deeper probing depth of the $E_{ph} = \SI{1486.6}{\electronvolt}$ Al-K$\alpha$ source (IMFP = \SI{28.9}{\angstrom}). The oxidation is accompanied by a steady increase in Cu 3p over several days (Fig. \ref{fig_4}(b)), greater than what was observed in AP-XPS. The link between oxidation and Cu migration is clear when looking at the trends in Figs. $2-4$: both Cu migration and \ce{Bi2O3} formation start right after dosing \ce{O2}, and when oxide growth slows between $1-4$ days, Cu growth also slows, changing only slightly. This suggests that surface oxides establish the conditions needed for Cu to diffuse towards the surface. 

There are several microscopic mechanisms that can promote Cu migration, such as a surface work function mismatch between \ce{Bi2O3} and \ce{Cu_xBi2Se3} that can drive the Cu to the surface with a built-in $E$-field. The work function difference between \ce{Bi2O3} and \ce{Bi2Se3} is estimated to be $\sim\SI{1.6}{\electronvolt}$ \cite{Morasch2013, Takane2016}, which would create a sufficiently large $E$-field near the surface. Cu is able to occupy five different sites in the Van der Waals gap and in interstitial vacancies \cite{Wang2011, Tumelero2016}, and some migration could be driven by Cu diffusion among these sites. Occupation of surface Se vacancies by Cu could also influence the Cu migration to the surface, which is possible due to the amphoteric character of Cu impurities in \ce{Bi2Se3} \cite{Vako1974}.

To connect the proposed changes in chemical composition to the measured XPS intensities, we model our experimental results with core-electron spectroscopy simulations using the National Institute of Standards and Technology (NIST) Simulation of Electron Spectra for Surface Analysis (SESSA) software/database \cite{Smekal2005, Werner2017}. SESSA can accurately simulate XPS spectra and peak intensities for different experimental conditions, geometries, and sample compositions using database reference values. As strongly suggested by our experimental observations, Cu and Se form a compositional gradient in the near-surface region of the material. A gradient structure consisting of several discretized, homogeneous layers with varying Cu, Bi, and Se compositions models a sample with Cu and Se gradients, shown in Figs. \ref{fig_5}(a,b). The initial Cu distribution is assumed to be constant. The gradients have a $X(Z)=Ae^{-Z/L}+B$ falloff when approaching the surface, with fitting parameters $A$, $L$, and $B$, which were chosen to match the boundary conditions observed in experiment. Due to the oxidation present after \ce{O2} dosing, the final simulated structure is also capped with a thin \SI{2}{\angstrom} overlayer of \ce{Bi2O3}. 
The red points in Figs. \ref{fig_5}(c,d) show the SESSA-calculated peak intensities for the initial and final structures, expressed in terms of the Cu/Bi and Se/Bi intensity ratios versus IMFP and the measurement depth. In the initial state, a structure with a single homogeneous Cu composition with no Cu gradient matches closely with the observed constant dependence versus IMFP in Fig. \ref{fig_2}(b). Initially, no Cu has migrated into the surface regions and no oxide has formed yet. In the final state after oxidation, the dual Cu, Se gradient structure agrees best with our measured XPS intensities in Figs. \ref{fig_5}(c,d), capturing both the increase in Cu and decrease in Se near the surface. We note that there is greater error when comparing the gradient to experiment at the smallest depth, due to the low photon flux at $E_{ph} = \SI{230}{\electronvolt}$.

Modelling the uppermost \SI{10}{\angstrom} of \ce{Cu_xBi2Se3} is motivated by the fact that the top \SI{10}{\angstrom} contributes predominately to the measured XPS intensity. Because of the exponential falloff at greater depths, compositional changes deeper in the sample are more difficult to detect. Additionally, \SI{10}{\angstrom} is the approximate thickness of one quintuple layer in \ce{Bi2Se3}, and the van der Waals gap can serve as a barrier to deeper oxidation of the material \cite{Green2016}. Thus, it is likely that the oxidation is limited to the uppermost quintuple layer in the AP-XPS experiment, with lesser contributions below.

There are other factors that are not present in this model which can also affect the evolution of \ce{Cu_xBi2Se3} surface chemistry. These include imperfect cleaves that can form step edge sites for oxidation \cite{Thomas2015}, lingering Cu remaining at the surface after the cleave, different Bi, Se surface terminations \cite{Biswas2015}, as well as nanosheet morphology \cite{Jia2016,Kong2011}. Other Cu, Se compositional distributions are possible and consistent with our experimental observations, such as ones with discontinuous step edges or sigmoidal distributions. 

\subsection{Conclusion}

In summary, we have observed an increase in the surface Cu content in the intercalated TI \ce{Cu_xBi2Se3} with ambient-pressure XPS measurements. Our results show that Cu migrates to the surface and is enhanced by the appearance and growth of surface oxides over several days, with the most pronounced changes seen in full atmosphere. Modelling Cu migration concomitant with Se depletion and oxidation matches our depth-selective XPS observations for a wide range of photon energies. 

These findings show that oxidation can be used as an approach for driving chemical species towards the surface of layered intercalated materials, and add additional chemical complexity that must be considered at TI surfaces exposed to ambient conditions. Chemically tailoring the surfaces of topological materials will be needed for realizing real-world environmental applications in chemical sensing, catalysis, and electronics. Most intriguingly, the proximity of the topological surface states to the observed chemical changes in \ce{Cu_xBi2Se3} points to further study of the effect of intercalants on the surface states of TIs, particularly on timescales that allow environmental changes to influence them. 

\vspace{5mm}

\begin{acknowledgments}
We thank Henrique Martins, Jonathan Denlinger, and Andrew Thron for helpful discussions. A.L.G. and S.N. acknowledge funding through the Laboratory Directed Research and Development (LDRD) Program at Lawrence Berkeley National Laboratory under a grant titled “Photoemission Investigations of Layered 2D Materials”, and A.L.G. acknowledges subsequent support from the Alfred P. Sloan Foundation (FG-2019-12170). L.J.F. acknowledges support of the Humboldt Foundation, Bonn, Germany. Purchase of the Kratos AXIS Supra+ XPS instrument used to collect data in Fig. 4 was supported by the National Science Foundation under the award NSF-MRI-1828238, and collection of these data were supported by NSF-DMR-1838532. ARPES experiments were supported by AFOSR Grant No. FA9550-18-1-0156. This research used resources of the Advanced Light Source, which is a US Department of Energy Office of Science User Facility under contract No. DE-AC02-05CH11231. R.R.U. and V.T. acknowledge support from the UC Lab Fees Research Program (LFR-20-653926) and the UC Davis Physics Liquid Helium Laboratory Fund.
\end{acknowledgments}

\bibliography{main_refs}

\end{document}

\title{Copper migration and surface oxidation of \ce{Cu_xBi2Se3} in ambient pressure environments: Supplementary Material}

\maketitle
%\vspace{5mm}

\section{XPS data analysis}

XPS intensities for Bi 5d and Se 3d are determined by fitting spectra to Voigt doublets with a Shirley background, using Kolibrik KolXPD for peak fitting and calculating peak areas. To determine oxide content, Bi oxide peaks were quantified by fitting a separate set of chemically shifted Voigt doublets to the spectra, keeping spin-orbit coupling and branching ratios fixed. To accurately quantify chemical content, all XPS intensities were weighed by the relative sensitivity factor (RSF) for each element at each excitation energy. The predominant term of the RSF is the photoionization cross-section $\frac{d\sigma}{d\Omega}$, calculated using elemental cross-sections and asymmetry values from the ELETTRA database \cite{Yeh1985,Yeh1993}, and converted to differential form by the Yeh and Lindau formula \cite{Yeh1985} using the experimental geometries for ALS Beamline 9.3.2 and for the Kratos AXIS SUPRA$^+$. The analyzer transmission function is not included in the RSF because the Cu 3p, Se 5d, and Bi 5d levels span a sufficiently small $\sim\SI{50}{\electronvolt}$ kinetic energy range, so the transmission function contribution is negligible. The photon flux and inelastic mean free path corrections were excluded as these are constant for each photon energy, and are not needed for calculating Cu/Bi and Se/Bi atomic ratios. In all figures, core level binding energies were offset in energy to be consistent with NIST and SESSA database values.

\section{\ce{Bi2Se3} loss feature correction}

\begin{figure*}[h]
\includegraphics[width=0.9\columnwidth]{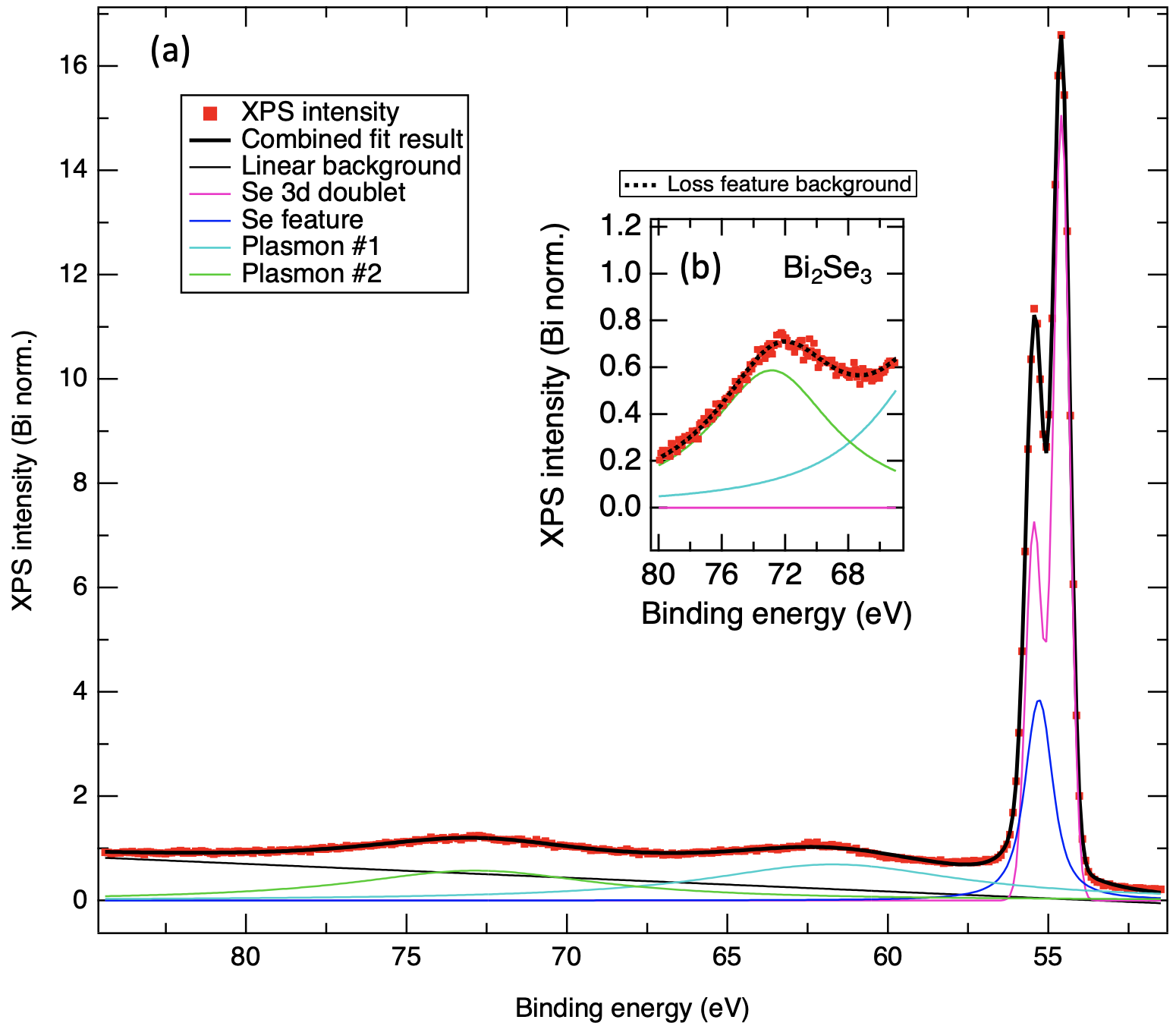}
\captionsetup{justification=centering,singlelinecheck=false}
	\caption{Loss feature fits in \ce{Bi2Se3} XPS spectra, right after cleaving. (a) Fitted XPS spectrum for \ce{Bi2Se3} including Se 3d core levels and plasmon loss features. The measured data is in red, the individual fitted peaks are the colored curves, and the combined fit result is the solid black curve. A linear background was fitted over the entire region. (b) Zoom in of plasmon loss feature background in \ce{Bi2Se3}, with the linear background subtracted.}
	\label{fig_Cu_loss}
\end{figure*}

This section describes the procedure that was implemented to systematically background-correct the AP-XPS \ce{Cu_xBi2Se3} spectra in the vicinity of the Cu 3p core levels to isolate the Cu 3p intensity (see main text). This procedure has the advantage of incorporating the plasmon energy loss features in the \ce{Cu_xBi2Se3} spectra, and is able to discriminate between the small Cu 3p intensity and the loss features for each excitation energy. The background is initially estimated by fitting to \ce{Bi2Se3} reference spectra at each excitation energy containing the plasmon energy loss features \cite{Nascimento1999}. To model the loss features, several Voigt peaks and a linear background were fit to the reference \ce{Bi2Se3} spectrum (Fig. \ref{fig_Cu_loss}(a,b)), forming the composite background that was subsequently fitted to the \ce{Cu_xBi2Se3} spectra in Fig. \ref{fig_Cu_loss}(c). The peak widths, peak ratios, and energy spacing between individual peaks were fixed, with only the amplitudes varying as a fitting parameter up to an overall binding energy offset determined by the Se 3d peak. The Cu 3p spectra were fitted to another Voigt doublet to quantify the XPS intensity of Cu 3p.

It is also possible to quantify Cu 3p after directly subtracting a normalized \ce{Bi2Se3} reference spectrum, which was done for a separate AP-XPS experiment in a mixed \ce{O2}/\ce{H2O} environment (not shown in the main text). A similar increasing Cu/Bi trend was seen with this background correction method, shown in Fig. \ref{fig_O2_H2O}(a). This will be discussed below in Section V.

\section{O 1s levels in \ce{Cu_{0.3}Bi2Se3}}

\begin{figure*}[h]
\includegraphics[width=1.0\columnwidth]{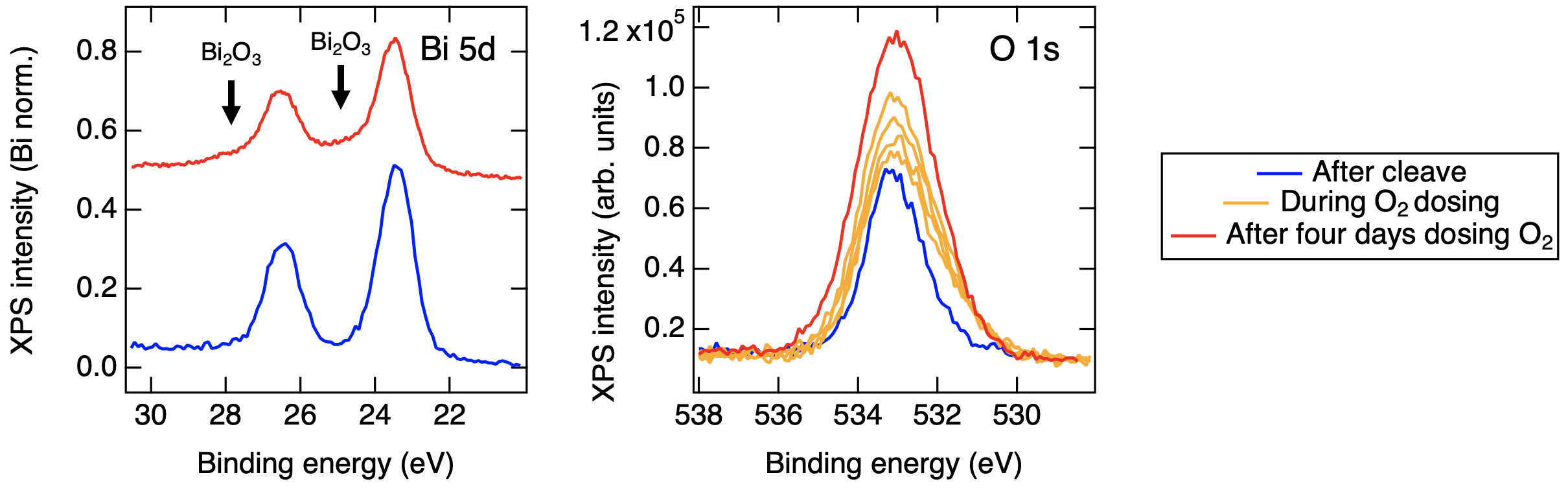}
\captionsetup{justification=centering,singlelinecheck=false}
	\caption{Oxidation of \ce{Cu_{0.3}Bi2Se3} in the AP-XPS experiment, taken with $E_{ph}=\SI{650}{\electronvolt}$. Left: growth of the Bi 5d levels initially after cleaving and after four days dosing \ce{O2}. Bi 5d peaks have been normalized to the total Bi intensity, and offset for clarity. Right: unnormalized O 1s peaks during the AP-XPS experiment.}
	\label{fig_oxides_650eV}
\end{figure*}

Additional spectra taken with $E_{ph}=\SI{650}{\electronvolt}$ in the AP-XPS experiment shows that the growth of the \ce{Bi2O3} features is coincident with the growth of the O 1s peak. Initially right after the cleave, adventitious O is adsorbed on the surface, as seen by the non-zero initial O 1s peak (Fig. \ref{fig_oxides_650eV}, right). After dosing \ce{O2}, the O 1s peak steadily grows, reaching the highest intensity at the end of the AP-XPS experiment when the Bi is most oxidized. We note that the O 1s peak is still growing at the end of the AP-XPS experiment, consistent with our longer timescale observations that show continued oxidation over several weeks.

\newpage
\section{Additional Cu migration data}

\begin{figure*}[h]
\includegraphics[width=1\columnwidth]{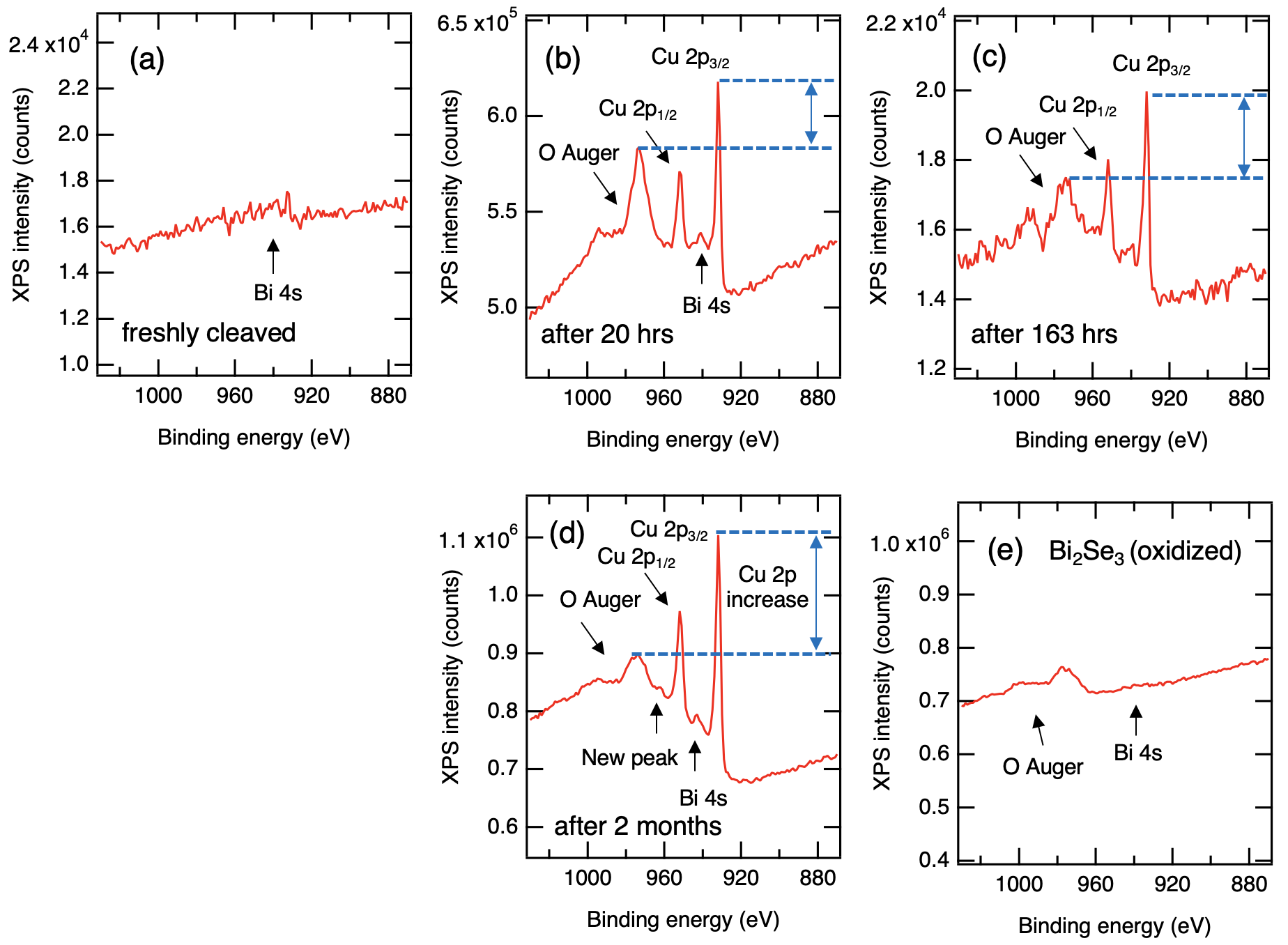}
\captionsetup{justification=centering,singlelinecheck=false}
	\caption{XPS spectra with Bi 4s and Cu 2p core levels, taken with $E_{ph} = \SI{1486.6}{\electronvolt}$. (a) XPS spectrum of a freshly cleaved \ce{Cu_{0.15}Bi2Se3} sample, showing a small Bi 4s peak and no visible Cu 2p peaks. (b) Spectrum of \ce{Cu_{0.15}Bi2Se3} after 20 hrs in air. Note the appearance of Cu 2p peaks, and O auger peaks. The distance between the O auger peak and the Cu 2p$_{3/2}$ peak is indicated by the spacing between the blue dashed lines. (c) Spectrum of \ce{Cu_{0.15}Bi2Se3} after 163 hrs, showing growth of the Cu 2p peaks compared to the O Auger peaks (longer blue arrow). (d) Spectrum of \ce{Cu_{0.15}Bi2Se3} after 2 months, showing further growth of the Cu 2p peaks. (e) Spectrum of \ce{Bi2Se3} sample kept in air for several months, for comparison.}
	\label{fig_Cu_2p}
\end{figure*}

At higher photon energy ($E_{ph} = \SI{1486.6}{\electronvolt}$) using the Kratos AXIS SUPRA$^+$, several higher binding energy Cu, Bi, and O levels are observable in the survey spectra: Cu 2p$_{3/2}$, Cu 2p$_{1/2}$ ($E_B=932.7,\SI{952.3}{\electronvolt}$), Bi 4s ($E_B= \SI{939}{\electronvolt}$), and O KL23L23, KL1L23 Auger peaks ($E_B=970.6,\SI{990.1}{\electronvolt}$). These are shown in Fig. \ref{fig_Cu_2p}. The Cu 2p levels near $E_B = 933-\SI{952}{\electronvolt}$ are more intense than Cu 3p and generally preferable for quantifying Cu and \ce{Cu2O}. However, tracking Cu 3p is needed over Cu 2p due to the strong overlapping Bi 4s level at $E_B = \SI{939}{\electronvolt}$, so only a qualitative analysis of the XPS spectra is possible. Nevertheless, separate observations also show that the Cu 2p levels increase in intensity after exposing \ce{Cu_{0.15}Bi2Se3} samples to air for several weeks, simultaneously with growing Cu 3p levels.

Freshly cleaved \ce{Cu_{0.15}Bi2Se3} shows only a small Bi 4s peak, with no Cu 2p or O Auger peaks. After \SI{20}{\hour} in air, Cu 2p peaks appear, consistent with the increase in the Cu 3p peaks at $E_B=\SI{76}{\electronvolt}$ after \SI{20}{\hour}. O Auger peaks also appear, which indicate surface oxidation. After \SI{163}{\hour}, the Cu 2p peaks continue to grow relative to the O Auger peak (blue arrows) demonstrating that Cu migration continues on the timescale as seen with Cu 3p. Interestingly, a small peak on the higher binding energy side of Cu 2p$_{1/2}$ forms after 2 months, which is not observed in a similarly oxidized \ce{Bi2Se3} sample. This peak is consistent with a CuO shakeup peak or a different oxidation state of Cu, indicating that a small amount of Cu probably oxidizes after many weeks of air exposure.

\newpage
\section{AP-XPS in a mixed \ce{O2}/\ce{H2O} environment}

\begin{figure*}[h]
\includegraphics[width=0.8\columnwidth]{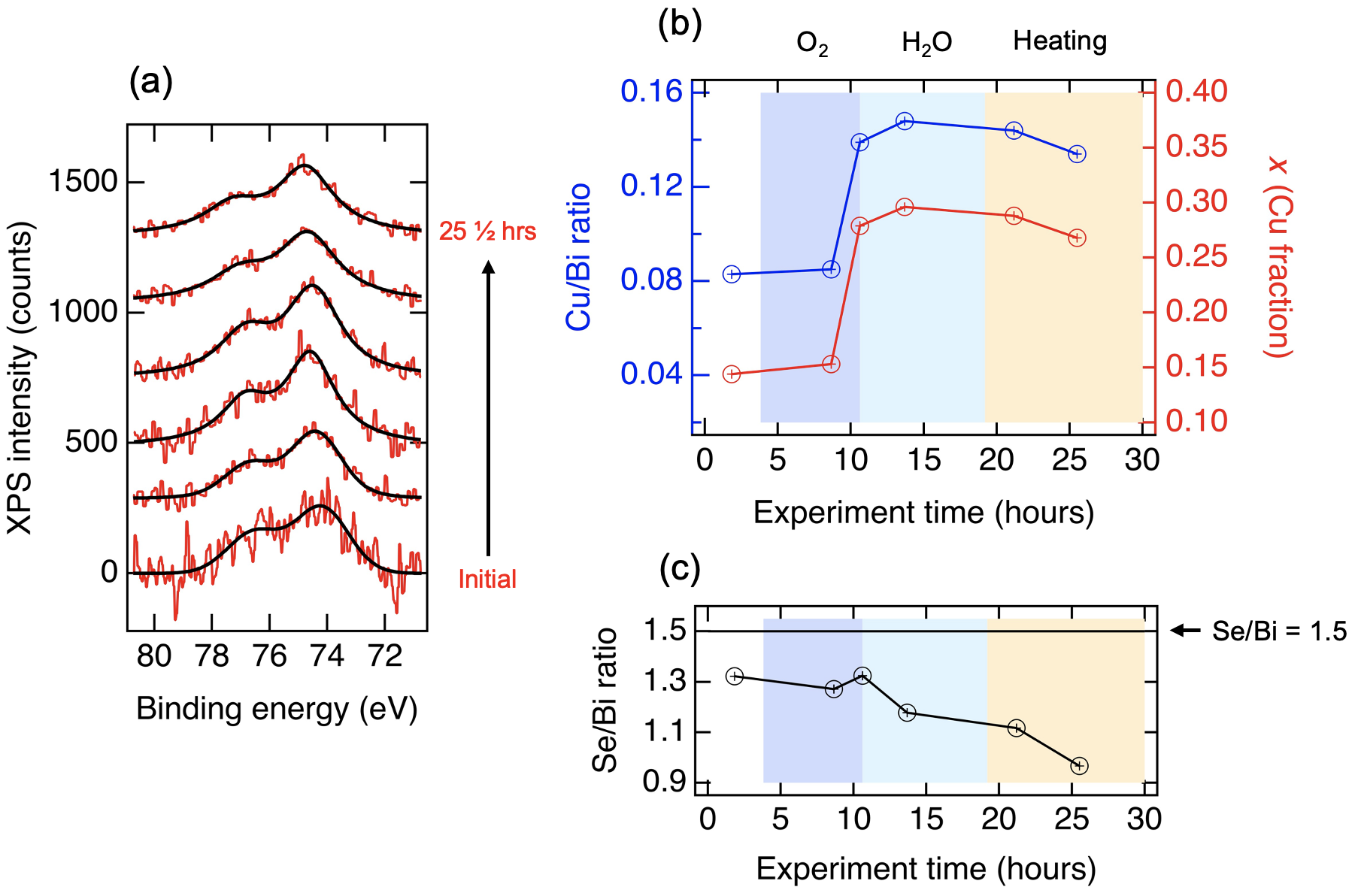}
\captionsetup{justification=centering,singlelinecheck=false}
	\caption{AP-XPS data for \ce{Cu_{0.15}Bi2Se3} in a mixed \ce{O2}/\ce{H2O} environment, measured with $E_{ph} = \SI{370}{\electronvolt}$. (a) Evolution of Cu 3p peaks, after correcting for the loss feature background. (b) Growth of Cu/Bi ratio and Cu fraction $x$ for several conditions: before dosing \ce{O2}, dosing \ce{O2}, dosing mixed \ce{O2}/\ce{H2O}, and heating when dosing \ce{O2}/\ce{H2O}. (c) Decrease of Se/Bi ratio for the same conditions.}
	\label{fig_O2_H2O}
\end{figure*}

Another set of AP-XPS experiments were performed in a mixed \ce{O2}/\ce{H2O} environment (100 mTorr/20 mTorr) showing a similar increase in Cu/Bi and decrease in Se/Bi consistent with the \ce{O2} experiment (Fig. \ref{fig_O2_H2O}). At \SI{10}{\hour} after measurements began, 20 mTorr \ce{H2O} was introduced into the chamber. At \SI{20}{\hour}, the sample was heated to \SI{80}{\celsius} from room temperature. The growth of Cu 3p peaks can expressed as loss feature-corrected spectra at different times in (Fig. \ref{fig_O2_H2O}(a)) and in terms of the atomic fraction Cu/Bi (Fig. \ref{fig_O2_H2O}(b)). The increase in signal-to-noise ratio in Fig. \ref{fig_O2_H2O}(a) is indicative of increased Cu content being measured at later times.

\begin{figure*}[h]
\includegraphics[width=1.0\columnwidth]{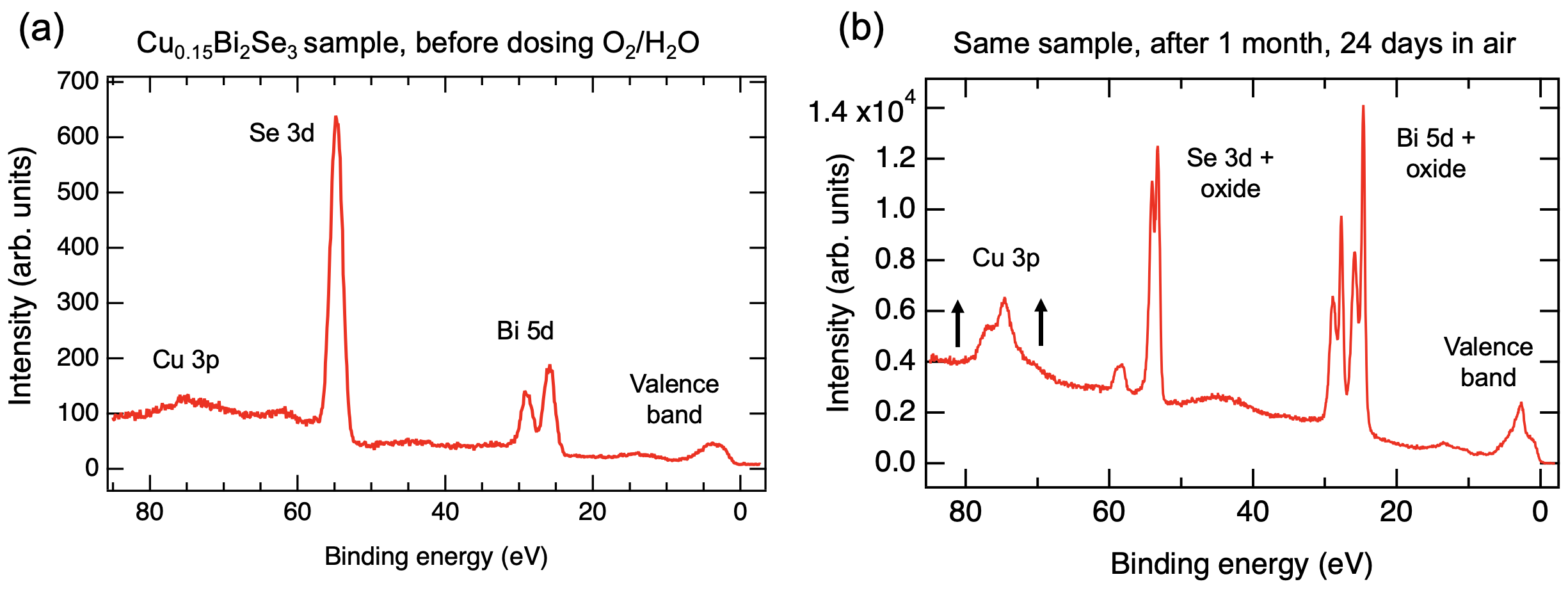}
\captionsetup{justification=centering,singlelinecheck=false}
	\caption{\ce{Cu_{0.15}Bi2Se3} sample after oxidation in air. (a) Shallow core levels including the valence band measured with $E_{ph} = \SI{900}{\electronvolt}$, prior to any \ce{O2}/\ce{H2O} dosing. (b) XPS spectrum taken after the AP-XPS experiment and 1 month, 24 days in air, using $E_{ph} = \SI{1486.6}{\electronvolt}$.}
	\label{fig_after_1_month}
\end{figure*}

The XPS spectrum of the same \ce{Cu_{0.15}Bi2Se3} sample was collected with the Kratos AXIS SUPRA$^+$ ($E_{ph} = \SI{1486.6}{\electronvolt}$) after 1 month and 24 days of aging in air. Shown in Fig. \ref{fig_after_1_month}, the Cu 3p peaks and the Se and Bi oxide peaks are considerably larger after extended air exposure. We note the growth of the valence band states in Fig. \ref{fig_after_1_month}(b), which are also consistent with greater Cu near the surface.

\newpage

\section{Se depletion and oxidation}

\begin{figure*}[h]
\includegraphics[width=0.7\columnwidth]{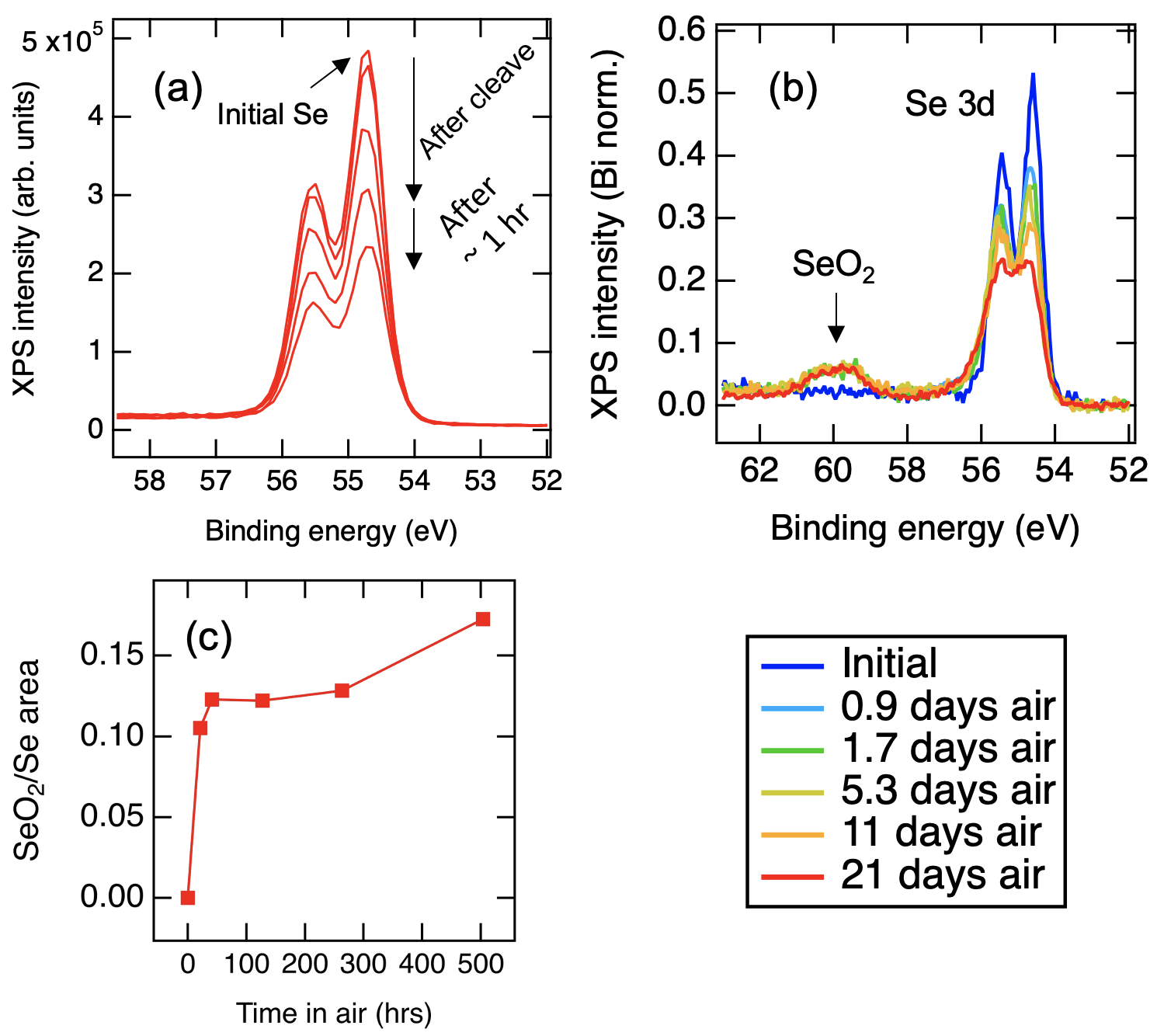}
\captionsetup{justification=centering,singlelinecheck=false}
	\caption{Se depletion and oxidation in \ce{Cu_{0.15}Bi2Se3}. (a) Se depletion after cleaving and dosing \ce{O2} in the AP-XPS experiment ($E_{ph} = \SI{370}{\electronvolt}$). (b) Se 3d peaks over several days of oxidation, with \ce{SeO2} visible at higher $E_B$ ($E_{ph} = \SI{1486.6}{\electronvolt}$). (c) Growth of \ce{SeO2} relative to \ce{Se} over several days.}
	\label{fig_Se_3d}
\end{figure*}

In the mixed \ce{O2}/\ce{H2O} AP-XPS experiments, the Se 3d peak intensities were observed dropping right after cleaving, and continue to decrease at later times (Fig. \ref{fig_Se_3d}(a)). These observations show that the Se content is highly volatile even in the earliest moments of handling the sample and when starting \ce{O2} dosing. Due to the lower partial pressure of \ce{O2}, no Se oxidation was observed in the AP-XPS experiments, and was not included in the SESSA model. Se oxidation can be observed, but only in measurements taken after oxidation in full atmosphere for several days. In Fig. \ref{fig_Se_3d}(b), a feature to the higher binding energy side of the Se 3d levels indicates that Se is oxidized, and follows the growth trend shown in Fig. \ref{fig_Se_3d}(c).

\clearpage

\section{\ce{Bi2Se3} and \ce{Cu_$\lowercase{x}$Bi2Se3} sample preparation}
\ce{Bi2Se3} samples were grown using a modified form of the Bridgman method \cite{NissonThesis2015,Bridgman1925}, starting with pure Bi and Se precursors mixed into an evacuated quartz ampoule in a 2:3 ratio. After melting of the initial mixture, the ampoule was placed into the furnace with a temperature gradient between \SI{750}{\celsius} - \SI{650}{\celsius}. The ampoule pulled through the gradient at a rate of 2-3 \SI{}{\milli\meter\per\hour} forming high purity \ce{Bi2Se3} at the end of the process. Additional information on the custom apparatus at UC Davis is available in the indicated reference \cite{NissonThesis2015}. A solution-based process was used to intercalate the \ce{Bi2Se3} with Cu to form \ce{Cu_xBi2Se3} \cite{Koski2012}. The Bridgman synthesized \ce{Bi2Se3} samples were placed in a tetrakisacetonitrile copper hexafluorophosphate solution and heated in 5 mL acetone at \SI{45}{\celsius}, just below reflux, for \SI{4}{\hour}. Samples were then heated under vacuum at \SI{180}{\celsius} for \SI{7.5}{\minute}, and the entire sequence was repeated for four cycles. 

After synthesis, samples were kept away from moisture during storage to avoid \ce{H2O} contamination. Samples were mounted on a conductive substrate with silver epoxy, and for cleaving in the AP-XPS experiment, ceramic top posts were adhered to the tops of the samples with silver epoxy. A mechanical arm was used to detach the top posts when cleaving \textit{in situ}.

\section{Characterization of superconductivity in \ce{Cu_xBi2Se3}}

\begin{figure}[h]
\includegraphics[]{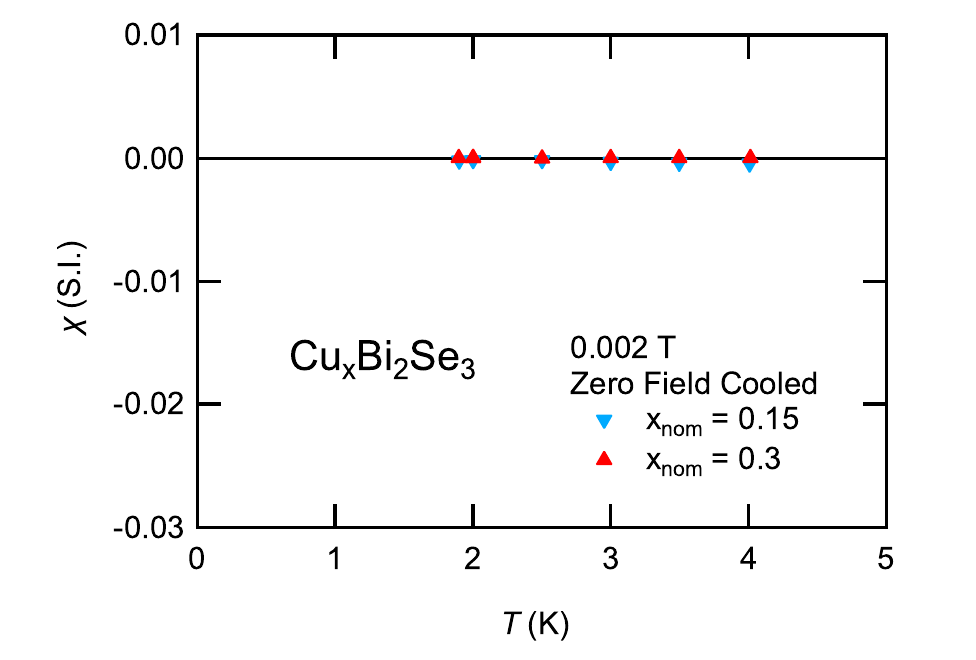}
\caption{\label{fig:Susceptibility} The DC magnetic susceptibility was measured on \ce{Cu_xBi2Se3} with nominal $x = 0.15$ and $x = 0.3$, down to 1.9\,K in a Quantum Design MPMS. No superconducting transition was observed in either sample.}
\end{figure}

We measured the DC magnetic susceptibility of a selection of \ce{Cu_xBi2Se3} samples in a Quantum Design SQUID Magnetometer (MPMS). The \ce{Cu_xBi2Se3} samples were wedged between multiple straws and oriented so the applied field was parallel to the c-axis. These choices were made to limit the diamagnetic background signal observed when alternative mounting methods are used. 
 
The $M/H$ vs. $T$ plot shows that none of the samples have a significant superconducting shielding fraction (less than 0.3\%). For a superconducting sample, one would expect 100\% shielding fraction (i.e. $M/H=-1$) in zero field cooled measurements. In the literature, samples have been reported to have shielding fractions up to 50\% for $x=0.4$ in Kriener et al. \cite{Kriener2011} and up to 56\% for Cu content of $x=0.35$ in Schneeloch et al. \cite{Schneeloch2015}. Additionally, a more recent study from Fang et al. \cite{Fang2020} reported a 17\% superconducting shielding fraction estimate after considering demagnetization effects.
The shielding fraction can depend on method of synthesis. Melt-growth methods, when quenching from above \SI{560}{\celsius} and annealing are known to produce superconducting samples, while floating zone crystal growth methods do not tend to produce superconducting samples \cite{Schneeloch2015}. Electrochemical synthesis also tends to promote superconductivity in \ce{Cu_xBi2Se3} \cite{Kriener2011}. 

\bibliography{supp_refs}